\documentclass[aps, pra,twocolumn,reprint,showpacs]{revtex4-1}%

\usepackage[latin1]{inputenc}
\usepackage[english]{babel}
\usepackage[T1]{fontenc}
\usepackage[dvips]{graphicx}
\usepackage{amsmath}
\usepackage{amsfonts}
\usepackage{amssymb}
\usepackage{epsfig}
\usepackage{graphicx,color}
\usepackage{dcolumn}

\newcommand{\eqnref}[1]{Eq.~(\ref{#1})}

\begin{document}
\title{Scaling of the explicit forms of the non-interacting kinetic-energy density functional}
\author{L\'{a}zaro~Calder\'{i}n}
\email{calderin@psu.edu}
\affiliation{Materials Research Institute and Research Computing and Cyberinfrastructure\\
             The Pennsylvania State University,\\
             University Park, State College, PA 16802, USA}

\date{\today}
\begin{abstract}
It has been previously proven that the Kohn-Sham kinetic energy functional scales homogeneously under 
generalized coordinate scaling.
Such scaling is observed by the von Weizs\"acker functional, 
but seems to be in contradiction with the scaling of the Thomas-Fermi functional. 
A puzzling situation, taking in to account that the von Weizs\"acker and Thomas-Fermi functionals are 
exact cases of the Kohn-Sham kinetic energy functional for one- or two-electron systems, 
and the non-interacting electron gas, respectively.
The apparent contradiction is resolved in this paper.
\end{abstract}
\pacs{31.15.E-, 71.15.Mb}
\maketitle

\section{Introduction}
%
In a recent paper we introduced a generalized homogeneous coordinate scaling of a set of Kohn-Sham (KS) wavefunctions $\phi_i(\mathbf{r})$ as \cite{PhysRevA.86.032510}
\begin{equation}
 \phi_{\alpha\beta m p}^{(i)}(\mathbf{r})=\alpha^{m/2} \phi_i(\beta^p
\mathbf{r}),
\label{eq:phiscaling}
\end{equation}
where $\alpha$ and $\beta$ are scaling positive real numbers, and the exponents $m$ and $p$ are real. 
Under such a general scaling the electron density scales as
\begin{equation}
 n_{\alpha\beta m p}(\mathbf{r})=\alpha^m n(\beta^p \mathbf{r});
\label{eq:nscaling}
\end{equation}
and we showed that the kinetic-energy density functional of a non-interacting systems of electrons of the same density, that is the KS kinetic energy functional $T_s[n]$, scales homogeneously as
\begin{equation}
T_s[n_{\alpha\beta m p}] = \frac{\alpha^m}{\beta^{p}} T_s[n].
\label{eq:tsscaling}
\end{equation}


The explicit general functional form of $T_s[n]$ it is not known. Instead, a Levy constrained search \cite{Levy85}, modified for consistency with the hamiltonian wave equation for a non interacting system \cite{PhysRevA.86.032510}, defines $T_s[n]$ as:
\begin{equation}
 T_s[n]=\min_{\{\phi\}_{\perp}\rightarrow n}-\frac{1}{2} \sum^{N_s}_{i=1} n_i \int \phi^*_i(\mathbf{r})
\nabla^2
\phi_i(\mathbf{r}) d^3r,
\label{eq:Ts}
\end{equation}
where $N_s$ are the number of states or wavefunctions, $n_i$ are the occupation number, and the search is performed over all the \textit{orthogonal sets of wavefunctions normalized to ${(\int d^3r\, n(\mathbf{r}) /N_e)^{1/2}}$}, $N_e$ being the number of electrons.

However, two particular kinetic energy explicit functionals of the density are known: the von Weizs\"acker \cite{vW35} and the Thomas-Fermi \cite{thomas,fermi} kinetic energy functionals. It can be readily proven that the von Weizs\"acker functional, $T_{vW}$, is exact for two electron systems; and it can be also shown that the Thomas-Fermi functional, $T_{TF}$, is exact for the non-interacting electron gas.

The analytical expression of $T_{vW}[n]$ is \cite{vW35}
\begin{equation}
 T_{vW}[n]=\frac{1}{8} \int d^3r \frac{|\nabla n(\mathbf{r})|^2}{n(\mathbf{r})};
\label{eq:tvw}
\end{equation}
and $T_{TF}[n]$ is given by \cite{thomas,fermi}
\begin{equation}
 T_{TF}[n] = C_{TF}\int d^3r\, n(\mathbf{r})^{5/3},
\label{eq:tf}
\end{equation}
where $C_{TF}=(3/10)(3\pi^2)^{2/3}$. The explicit analytical exact forms allow us to study their compliance with the homogeneous general scaling property of $T_s[n]$; and at the same time verify the general scaling property.

One immediately can prove that $T_{vW}[n]$ does obey \eqnref{eq:tsscaling}; but $T_{TF}$ does not seem to follow it.
In fact, a direct substitution of the scaled density into the $T_{TF}$ functional reveals that 
\begin{equation}
T_{TF}[n_{\alpha\beta m p}] = \frac{\alpha^{5m/3}}{\beta^{3p}} T_{TF}[n],
\label{eq:tfscaling}
\end{equation}
which is in complete contradiction with \eqnref{eq:tsscaling}, given that in the limit of the uniform electron density, for which the TF functional is supposed to yield the exact kinetic energy $T_s$ for all densities, the TF functional delivers the wrong result for the scaled uniform electron density.
That is, TF does not scale properly even in the case when it is exact.

Therefore, something must be wrong either with the general scaling result for $T_s$ (\eqnref{eq:tsscaling}), or with the Thomas-Fermi functional. The general scaling property of $T_s[n]$ was proved in different ways, and extensively tested numerically \cite{PhysRevA.86.032510}. Besides, $T_{vW}$ also obeys the homogeneous scaling property of $T_s[n]$. Hence, what is in order is a careful review of the construction of the Thomas-Fermi functional, starting with the scaling of the wavefunctions, which is the goal of this paper.

We start with a step-by-step review of the electron gas and the definition of the Thomas-Fermi functional, and proceed to use the same steps to analyze the electron gas for the scaled wavefunctions and its corresponding Thomas-Fermi functional.


\section{Free electron gas}
A gas of $N_e$ non-interacting electrons confined to a box of size $a$, and under Born-Karman periodic boundary conditions, has states described by the plane waves 
\begin{equation}
 \phi_{\mathbf{k}}(\mathbf{r})=\frac{1}{a^{3/2}} e^{ i \mathbf{k}\cdot\mathbf{r}},
\end{equation}
with discrete wave-vectors
\begin{equation}
\mathbf{k}= \frac{2\pi}{a} (n_x \hat{i} + n_y \hat{j} + n_z \hat{k} ),  
\end{equation}
defined on cartesian coordinates, and $n_x, n_y$ and $n_z$ taking positive and negative integer values. The electron density is given by
\begin{equation}
n(\mathbf{r}) = \sum_\mathbf{k} n(\mathbf{k}) \phi^*_\mathbf{k}(\mathbf{r}) \phi_\mathbf{k}(\mathbf{r});
\end{equation}
where $n(\mathbf{k})$ are occupation numbers that take values $0,1$ or $2$. The sum is done over all the $\mathbf{k}$ vectors and it is straight forward to get for the density
\begin{equation}
 n(\mathbf{r}) = \frac{N_e}{V}\equiv \bar n,
 \label{eq:density}
\end{equation}
where $V=a^3$ is the volume of the box,
given that
\begin{equation}
\sum_\mathbf{k} n(\mathbf{k})=N_e.
\end{equation}
The total energy of the non-interacting gas is purely kinetic and it is readily calculated from
\begin{equation}
T_g[\bar n] = -\frac{1}{2} \sum_\mathbf{k} n(\mathbf{k}) \int d^3r\, \phi^*_\mathbf{k}(\mathbf{r}) \nabla^2 \phi_\mathbf{k}(\mathbf{r}),
\end{equation}
as
\begin{equation}
T_g[\bar n] = \frac{1}{2} \sum_\mathbf{k} n(\mathbf{k})\, k^2.
\end{equation}

The next step towards the construction of the Thomas-Fermi functional is to take the limit to the continuum in $\mathbf{k}$-space by allowing the size of the box and the number of electrons to increase infinitely, but keeping the density constant. In such a limit we have that
\begin{equation}
\sum_{\mathbf{k}} \rightarrow \frac{V}{(2\pi)^3} \int d^3k, 
\end{equation}
where the integral is taken over all the $k$-space and therefore the density takes the form
\begin{equation}
\bar n =  \frac{V}{(2\pi)^3} \int d^3k\,\, n(\mathbf{k})\, \phi^*_\mathbf{k}(\mathbf{r}) \phi_\mathbf{k}(\mathbf{r});
\end{equation}
and the kinetic energy density per unit of volume $t_g[\bar n]$ is calculated by
\begin{equation}
t_g[\bar n] = \frac{1}{2(2\pi)^3}  \int d^3k\,\, n(\mathbf{k})\, k^2.
\end{equation}

The electrons of the same $|\mathbf{k}|$ have the same energy and all of them will occupy the states inside an sphere
of radius $k_F$ (Fermi vector). We can then define the occupation number by an step function which is two inside the Fermi's sphere and zero out; allowing to write $t_g$ as
\begin{equation}
t_g[\bar n] = \frac{1}{10\pi^2} k_F^{5}. 
\end{equation}
and the density as
\begin{equation}
\bar n =  \frac{1}{3\pi^2} k_F^3,
\end{equation}
from which we get
\begin{equation}
k_F = ( 3\pi^2 \bar n )^{1/3}.
\label{eq:kf}
\end{equation}

Finally the kinetic energy per unit of volume can be written in terms of the electron density as
\begin{equation}
t_g[\bar n] = C_{TF}\, \bar n ^{5/3},
\label{eq:tg_of_n}
\end{equation}
using \eqnref{eq:kf}.

\section{Thomas-Fermi functional}
%
The Thomas-Fermi functional of an isolated system of density $n(\mathbf{r})$ is defined as
\begin{equation}
 T_{TF}[n] = \int d^3\mathbf{r}\, \left.{t_g[\bar n]}\right|_{\bar n = n(\mathbf{r}) },
\label{eq:tf_def}
\end{equation}
where the integral is taken over all the space. A direct substitution of \eqnref{eq:tg_of_n} into \eqnref{eq:tf_def} yields the expression for the TF kinetic energy functional (\eqnref{eq:tf}).

Once we have obtained the expression of $T_{TF}$ in terms of the density one just substitutes the 
electron density  under the integral to get the kinetic energy. But notice all what is behind that evaluation:
{\it First, for each position in the space the value of the density $n(\mathbf{r})$ is calculated and an electron gas of the same density $\bar n\equiv n(\mathbf{r})$
is created, characterized by plane waves and its kinetic energy density calculated. Second, the integral 
over the space of that kinetic energy density as a function of the position is taken to get the kinetic energy. 
}
Of course, repeating this process for each density will lead to the
same result as the one obtained by evaluating the TF functional at the electron density of interest.
However, as we will see next, that is not the case for the 
scaled density.

\section{Thomas-Fermi for the scaled density}
%
We proceed to scale the plane waves according to \eqnref{eq:phiscaling}; that is
\begin{equation}
 \phi^{(\mathbf{k})}_{\alpha\beta m p}(\mathbf{r})=\alpha^{m/2}\phi_{\mathbf{k}}(\beta^p\mathbf{r})
=\frac{\alpha^{m/2}}{a^{3/2}} e^{ i \beta^p \mathbf{k}\cdot\mathbf{r}};
\end{equation}
which comply with the Born-Karman boundary conditions in a box of size $a/\beta^p$, and together with \eqnref{eq:density} yields the scaled electron density
\begin{align}
n_{\alpha\beta m p}(\mathbf{r})
=& \alpha^m \sum_\mathbf{k} n(\mathbf{k}) \phi^{(\mathbf{k})*}(\mathbf{r}) \phi^{(\mathbf{k})}(\mathbf{r})
=\alpha^m {\bar n}\equiv {\bar n }_{\alpha\beta m p};
\end{align}
while the kinetic energy for the scaled density in terms of the wavefunctions
\begin{align}
T_g[\bar n_{\alpha\beta m p}] 
=& -\frac{1}{2} \sum_\mathbf{k} n(\mathbf{k}) 
\int d^3r\, \phi^{(\mathbf{k})*}_{\alpha\beta m p}(\mathbf{r}) \nabla^2 \phi^{(\mathbf{k})}_{\alpha\beta m p}(\mathbf{r}),
\end{align}
takes the form
\begin{align}
T_g[\bar n_{\alpha\beta m p}]=& \frac{\alpha^m\beta^{2p}}{2 a^3}\sum_\mathbf{k}n(\mathbf{k}) k^2
\int_0^{a/\beta^p}\int_0^{a/\beta^p}\int_0^{a/\beta^p} dx dy dz;
\end{align}
which by \eqnref{eq:tg_of_n} is reduced to
\begin{align}
T_g[\bar n_{\alpha\beta m p}]=& \frac{\alpha^m}{\beta^{p}} T_g[\bar n].
\end{align}

The next step is to take the limit to the continuum in $\mathbf{k}$-space, as in the case of the unscaled density, limit that yields for the kinetic energy density per unit of volume
\begin{equation}
t_g[\bar n_{\alpha\beta m p}] = \frac{1}{2(2\pi)^3} \frac{\alpha^m}{\beta^{p}} \int d^3k\,\, n(\mathbf{k})\, k^2;
\end{equation}
which using again \eqnref{eq:tg_of_n} can be written as
\begin{equation}
t_g[\bar n_{\alpha\beta m p}] = \frac{\alpha^m}{\beta^{p}} t_g[\bar n].
\label{eq:tg_of_n_scaling}
\end{equation}


Therefore, the Thomas-Fermi functional for the scaled density
\begin{equation}
 T_{TF}[n_{\alpha\beta m p}] = \int d^3\mathbf{r}\, \left.{t_g[\bar n_{\alpha\beta m p}]}\right|_{\bar n_{\alpha\beta m p} = n_{\alpha\beta m p}(\mathbf{r}) },
\end{equation}
yields, following \eqnref{eq:tg_of_n_scaling},
\begin{equation}
 T_{TF}[n_{\alpha\beta m p}] = \frac{\alpha^m}{\beta^p}\int d^3\mathbf{r}\, \left. t_g[\bar n]\right |_{\bar n = n(\mathbf{r}) };
\end{equation}
and hence, according to \eqnref{eq:tf_def} we can write
\begin{equation}
 T_{TF}[n_{\alpha\beta m p}] = \frac{\alpha^m}{\beta^p} T_{TF}[n];
\end{equation}
which is in complete agreement with the general homogeneous scaling properties of the KS kinetic energy functional $T_s[n]$ (\eqnref{eq:tsscaling}).

In conclusion, we have shown that both the von Weizs\"acker and the Thomas-Fermi kinetic energy functionals scale according the the general scaling properties of the Kohn-Sham kinetic energy functional under generalized homogeneous coordinate scaling.

\textbf{Acknowledgments:} Discussions with Malcolm J. Stott, John Perdew and Ismaila Dabo are gratefully acknowledged.


\bibliographystyle{apsrev4-1}
\bibliography{bib}

\end{document}